\documentclass[pre,preprint,showkeys,amsmath,amssymb,superscriptaddress]{revtex4}

\usepackage[dvips]{graphicx}
\usepackage{dcolumn}
\usepackage{bm}
\usepackage[titletoc,title]{appendix}
\usepackage{longtable}
\usepackage{bm}
\usepackage{color}
\usepackage[section]{placeins}

\begin{document}
\setlength\LTcapwidth{6.5in}
\setlength\LTleft{0pt}
\setlength\LTright{0pt}

\title{Microcanonical-ensemble computer simulation of the high-temperature
expansion coefficients of the Helmholtz free-energy of a Square-well fluid}

\author{Francisco Sastre}
\email{sastre@fisica.ugto.mx}
\affiliation{%
Departamento de Ingenier\'ia F\'isica,\ Divisi\'on de Ciencias e Ingenier\'ias,\\
Campus Le\'on de la Universidad de Guanajuato%
}

\author{Elizabeth Moreno-Hilario}
\email{morenohe2013@licifug.ugto.mx}
\affiliation{%
Divisi\'on de Ciencias e Ingenier\'ias,\\
Campus Le\'on de la Universidad de Guanajuato%
}

\author{Maria Guadalupe Sotelo-Serna}
\email{sotelosm2013@licifug.ugto.mx}
\affiliation{%
Divisi\'on de Ciencias e Ingenier\'ias,\\
Campus Le\'on de la Universidad de Guanajuato%
}

\author{Alejandro Gil-Villegas}
\email{gil@fisica.ugto.mx}
\affiliation{%
Departamento de Ingenier\'ia F\'isica,\ Divisi\'on de Ciencias e Ingenier\'ias,\\
Campus Le\'on de la Universidad de Guanajuato%
}

\date{\today}

\begin{abstract}

The Microcanonical Ensemble computer simulation method (MCE)
is used to evaluate the perturbation terms $A_i$ of the Helmholtz free energy of a Square-Well (SW) fluid.
The MCE method offers a very efficient and accurate procedure for the determination
of perturbation terms of discrete-potential systems such as the SW fluid and
surpass the standard NVT Canonical Ensemble Monte Carlo method,
allowing the calculation of the first six expansion terms. Results are presented for the case of
a SW potential with attractive ranges $1.1 \le \lambda \le 1.8$.
Using semiempirical representation of the MCE values for 
$A_i$, we also discuss the accuracy in the determination of the phase diagram of this system.
\end{abstract}



\maketitle

\section{Introduction}


Discrete potential (DP) systems have been investigated
over the years in order to model properties of real substances,
and offer a general and useful
approach to represent properties of model systems interacting via continuous
potentials~\cite{Cummings84}-\cite{alb:jcp10}.
The discretization process 
is based on the square-well (SW) fluid system,
that becomes the basic input to describe more general discrete potentials and
their mixtures. Examples of applications of DP systems
to model real fluids have been biodiesel blends~\cite{perdomo1,perdomo2} and
two-dimensional equations of state for adsorption isotherms of noble gases~\cite{delrio91},
asphaltenes
confined in porous materials~\cite{Castro09}, as well as carbon dioxide,
hydrogen, water and methanol adsorbed onto 
carbon-based substrates~\cite{Jimenez2008,Trejos14,Trejos17}. 

The SW pair potential for spherical particles of diameter $\sigma$ separated
by a distance $r$ is given by
\begin{equation}
\phi(r)=\left\{
\begin{array}{ccc}
\infty & \mbox{if} & r\leq \sigma \\
\setlength\LTright{0pt}
\setlength\LTright{0pt}
\epsilon & \mbox{if} & \sigma < r \leq \lambda\sigma \\
0 & \mbox{if} &  r > \lambda\sigma \\
\end{array}
\right.,\label{potencial}
\end{equation}
where $\epsilon$ is the depth-well energy and $\lambda$ is the range of the attractive interaction. 

This system has been studied in great detail since it is a simple but at the same time
complete model in order to understand the effect 
of non-conformal changes in the phase diagram and the dependence of the phase stability
on the range of the attractive forces~\cite{Frenkel1994}-\cite{ben:jcp05}. At the same time,
it has been a key model for assessing the role played by
multipolar moments when they are explicitly taken into account in the
thermodynamic modeling of DP systems~\cite{guevara94}-\cite{ben:jcp11}.
Thermodynamic and structural properties
of the SW system have been fully determined over the years through computer simulations ,
thermodynamic perturbation methods and integral equation
theories~\cite{Smith1970}-\cite{Zhou2013}.
The combined use of these results has allowed to
obtain  accurate equations of state valid for the whole range of the fluid phase diagram, 
and in this way the SW system has become a key ingredient
on robust approaches, such as the SAFT-VR method, to model a great variety of 
three and two-dimensional chain and branched molecular 
systems~\cite{gil:jcp97,galindo98,galindo05,Mccabe99,gil07}, including the description
of lamellar phases~\cite{Jimenez13}. 
Besides the SAFT models, alternative approaches to model anisotropic systems have
taken into account the non-spherical shape of the molecules combined with a SW interaction
~\cite{guevara99}-\cite{Jackson14}.
More recently, it has been possible to generate 
accurate quantum thermodynamic
perturbation theories based on the SW model ~\cite{Trejos2012,Serna2016}.

In spite of these advances in the characterization of the SW system, and consequently
in the improvement on the accuracy of equations of state for DP fluids,
there are still challenges to assume. For example, the caloric properties as
given by heat capacities and adsorption heats, require a better description 
of higher order perturbation terms.  
Following the Zwanzig's high-temperature perturbation expansion (HTE)~\cite{Zwanzig54},
the excess Helmholtz free energy
for a SW fluid can be expressed as a series on powers of the inverse thermal energy $\beta = 1/kT$,
\begin{equation}
\frac{A^E}{NkT}= \frac{A_0^E}{NkT}+\sum_{n=1}^{\infty} \beta^n A_n,
\end{equation}
where $A_n$ are the perturbation terms given by~\cite{Alder1972}
\begin{equation}
A_n=\lim_{\beta\to 0}\frac{1}{n!} \frac{\partial^n (A/NkT)}{\partial\beta^n}\label{coeficientes}
\end{equation}
and $A_0^E$ is the hard-spheres (HS) Helmholtz free energy.
The standard method to obtain exact values of the perturbation terms
is through NVT-ensemble Monte Carlo simulations ~\cite{Alder1972}, where
the perturbation terms are given as statistical averages of the attractive energy and
their fluctuations, obtained from HS configurations. Recently, 
the Microcanonical computer simulation method (MCE) \cite{Sastre2015}
was proposed as an
straightforward approach to simulate thermodynamic properties of the SW fluid.
Since in this ensemble the inverse thermal energy $\beta$ is given
in terms of the internal energy, it is possible to have 
a direct determination of   
the perturbation terms $A_n$ according to the Zwanzig expansion. As we will discuss here,
the MCE method can be used to accurately obtain HTE coefficients of order as higher as six,
a task difficult to accomplish using the  
standard NVC ensemble MC method.
The description of the method is given in Section II and its  application to the case of the SW
potential is presented in Section III, where parametric expressions for $A_n$ are given in order
to generate an equation of state, that is used for the evaluation of 
internal energies and isochoric heat capacities.
Finally, in Section IV the main conclusions of this work are given.

\noindent

\section{Simulation Method}

\begin{figure*}[!ht]
\begin{center}
\includegraphics[width=15.0cm,clip]{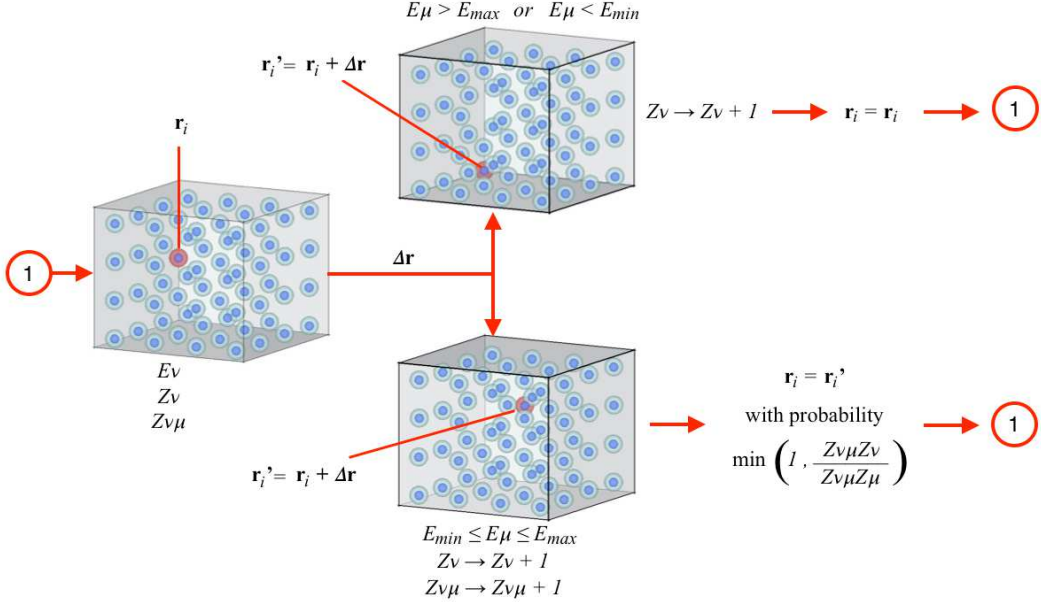}
\caption{\label{ejemplo} (Color on line)
Schematic representation of the 
Microcanonical Ensemble computer simulation method.}
\end{center}
\end{figure*}

We describe briefly the simulation MCE method applied to fluids,
that has been presented in our previous work~\cite{Sastre2015},
where the reader can found more details.
This method 
was originally developed for the
computer simulation of the Ising model using a cluster algorithm~\cite{Huller2002}
that optimizes a Broad Histogram Method~\cite{Oliveira1996}-\cite{Kastner2000}.
Since the SW fluid has a discrete set of energy values as in the Ising model, the 
cluster algorithm is easily transferable to the SW fluid.

We consider a
system of $N$ spherical particles of diameter $\sigma$
contained in a volume $V$ that interact via the
SW pair potential given by~Eq.(\ref{potencial}).
The system can be characterized by its
available energy levels $E_\nu$, 
that are multiples of $\epsilon$, 
\begin{equation}
E_\nu = -\nu\epsilon = \sum_{k,l\neq k} \phi(r_{kl}),
\end{equation}
where $r_{kl}$ is the distance between
the centers of the $k$ and $l$ particles
and $\nu$ is the number of particles pair
that satisfies $\sigma < r_{kl} \leq
\lambda\sigma$. The number of microstates that share
the same energy $E_\nu$ is denoted by
$\Omega(E_\nu)$. When a particle in a given
microstate is displaced by $\Delta\mathbf{r}$,
a new microstate is generated. After the displacement
is applied to all $N$ particles for each $\Omega(E_\nu)$
microstates, then $N\Omega(E_\nu)$ new microstates are created and 
just a given number of them, denoted by $V_{\nu\mu}$,  will have energy $E_\mu$. 
By the condition of reversibility,  
$V_{\nu\mu}=V_{\mu\nu}$ ~\cite{Poliveira1998}.
Two different protocols of allowed moves are used, based on 
either reversibility or equiprobability.
Thus if there is a random selection of one of the
$\Omega(E_\nu)$ microstates with energy $E_\nu$ and a
particle that is displaced by
$\Delta \mathbf{r}$, 
then the probability that the system reaches the energy $E_\mu$ will be given as 
\begin{equation}
P(E_\nu \to E_\mu)=\frac{V_{\nu\mu}}{N\Omega(E_\nu)}\label{probaij},
\end{equation}
and for the reversal case 
\begin{equation}
P(E_\mu \to E_\nu)=\frac{V_{\mu\nu}}{N\Omega(E_\mu)}\label{probaji}.
\end{equation}
In the simulation these probabilities are obtained by
calculating the rate of attempts $T_{\nu\mu}$ to go from 
level $\nu$ to level $\mu$.
This can be achieved with two quantities:
\begin{itemize}
\item The number of times that the system spends in 
level $\nu$, denoted by $z_\nu$.
\item The number of times that the system attempts to go from level $\nu$
to level $\mu$, denoted by $z_{\nu\mu}$.
\end{itemize}

The values $z_\nu$ and $z_{\nu\mu}$ are obtained as follows:
\begin{enumerate}
\item With $\nu$ as the initial state, a particle is randomly chosen
and $z_\nu$ is {\bf always} incremented by 1.
\item Then a new energy $E_\mu$ is evaluated if a random displacement is applied to
the chosen particle.
\item If $E_\mu$ is an allowed energy level, 
then $z_{\nu\mu}$ is {\bf always} incremented by 1, independently
of the acceptation of the particle displacement. 
It is possible to
restrict the energy levels by 
discarding all cases where $E_\mu<E_{min}$ or $E_\mu> E_{max}$.
\item The particle displacement is accepted if $T_{\nu\mu}<T_{\mu\nu}$, otherwise 
the displacement is chosen with a probability equal to $T_{\mu\nu}/T_{\nu\mu}$. 
\end{enumerate}
The last condition assures that all levels are visited with equal probability,
independently of their degeneracy.
The initial values for $z_i$ and $z_{\nu\mu}$ can be any positive number and 
after a large number of particle displacement attempts it will be observed that
\begin{equation}
\frac{z_{\nu\mu}}{z_\nu}\to T_{\nu\mu},
\end{equation}
and consequently
\begin{equation}
\frac{T_{\nu\mu}}{T_{\mu\nu}}=\frac{\Omega(E_\mu)}{\Omega(E_\nu)}.\label{dos}
\end{equation}
This algorithm is highly efficient for evaluating the ratios $\Omega(E_\nu)/\Omega(E_\mu)$,
or the differences $S_\nu-S_\mu$, using the microcanonical ensemble relation $S(E)=k\ln{[\Omega(E)]}$,
because the number of times that the random number generator
is required during the simulation is smaller than 
for NVT simulations.
The
efficiency increases if the number of allowed levels $(E_{max}-E_{min})/\epsilon$ is decreased.
The entropy can be expressed as
\begin{equation}
S(E_\mu)=S(E_\nu)+\epsilon\eta \beta(E_\nu)+\ldots,
\end{equation}
where $\beta(E)=\partial S/\partial E$ is the inverse thermal energy, and $\eta$
is an integer such that $E_\mu=E_\nu+\eta\epsilon$. The rest of the terms can be discarded, as long as $N$ is
large enough, and then we have
\begin{equation}
\ln(T_{\nu\mu}/T_{\mu\nu})\approx \frac{\epsilon\eta}{k} ~
\left.\frac{\partial S}{\partial E}\right|_\nu. \label{sol_beta}
\end{equation}
The last equation allow us to evaluate
the inverse thermal energy as function of the internal energy ~\cite{Sastre2015},
\begin{equation}
\beta(E_\nu)=\frac{1}{6}\sum_{\eta=-3}^{3} \frac{1}{\eta}\ln(T_{\nu,\nu+\eta}/T_{\nu+\eta,\nu}),~~\eta\neq 0,
\end{equation}
as well as the isochoric heat capacity $c(E)$,
\begin{equation}
c(E) = -\frac{[\beta(E)]^2}{S^{''}},
\end{equation}
where $S^{''} = \partial \beta/\partial E$.
In Figure \ref{ejemplo} we summarize the MCE method.

The very relevant feature of the MCE method
to determine the HTE coefficients $A_n$ is given by the fact
that when the entropy is
maximum,  
the curves $\beta(E)$
can be accurately evaluated 
around the values of energy that corresponds to $\beta=0$ or equivalently to
the limit $T\to\infty$, that is precisely
the limit where the HTE coefficients $A_n$ are defined, starting with
the mean attractive energy,
\begin{equation}
A_1= \lim_{\beta^*\to 0} u^*,
\end{equation}
where $u^*=E/N\epsilon$ and $\beta^*=\epsilon\beta$ are the reduced energy and inverse thermal energy, respectively.
The higher-order perturbation terms can be obtained according to the expression
\begin{equation}
A_n= \frac{1}{n!} \lim_{\beta^* \to 0} \frac{\partial^{(n-1)}u^*}{\partial \beta^{*(n-1)}} \label{formulaAs}
\end{equation}

\section{Results}

Computer simulations were performed for SW systems comprised
of $N=512$ particles contained in a unitary box
with periodic boundary conditions
and reduced densities $\rho^* = \rho\sigma^3$ between $0.1$ and $0.7$.
SW attractive ranges were taken within
the interval $1.1 \le \lambda \le 1.8$,
that are the values required to model real substances,
as dos Ramos {\it et al.} have determined for a wide range of systems 
in the context of the
SAFT-VR approach~\cite{galindo09}.
Restricting simulations to energy levels where $|\beta^*|\le 0.1$,
the corresponding values of the inverse thermal energy as function of the internal energy $u^*$
were represented by 
a second-order degree polynomial, for fixed values of $\rho^*$ and $\lambda$,
\begin{equation}
\beta^*= a_0+a_1 u^* + a_2 {u^*}^2 \label{polinomio}.
\end{equation}
Since the evaluation of the $A_n$ coefficients is performed
under the condition of a maximum value of the entropy, or equivalently
$\ln\Omega(u^*)$, 
then 
a quadratic expression for $\beta^*$ as a function of $u^*$ 
is justified due to the gaussian behavior 
of
$\Omega(u^*)$
around $\beta^* =0$. 

\begin{figure}
\begin{center}
\includegraphics[width=8.50cm,clip]{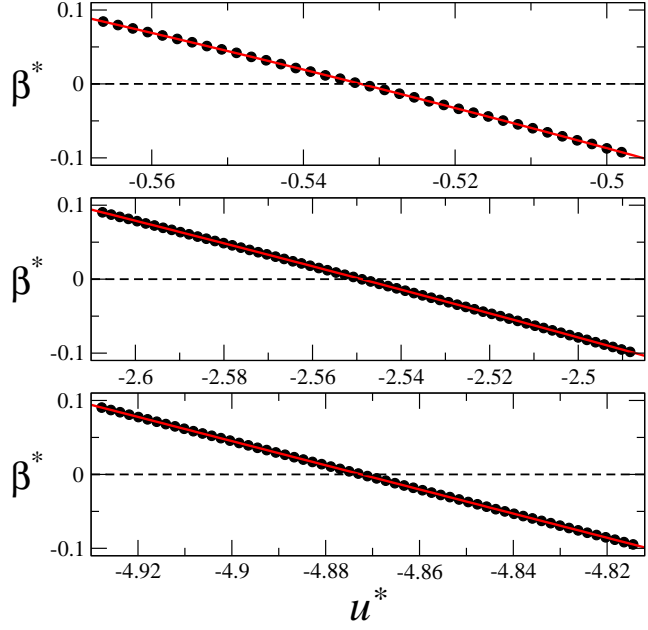}
\caption{\label{ejemplo2} (Color on line)
Inverse temperature as a function of the internal energy $u^*$ for the SW fluid with range
$\lambda=1.5$ and densities $\rho^*=0.1, 0.4$ and $0.7$, from top to bottom.
Black dots denote MCE simulation results and the red solid line is the second-order polynomial fit
given in Eq.(15).
The value of $u*$ for $\beta = 0$ corresponds to the first perturbation term $A_1$.
}
\end{center}
\end{figure}

In Figure \ref{ejemplo2} the case $\lambda=1.5$ is presented for densities
$\rho^* = 0.1, 0.4$ and $0.7$ from top to bottom. Notice that the mean attractive energy
$A_1$ is given as $u^*$ for $\beta = 0$.

Equation (15) can be easily inverted as
\begin{equation}
u^*= -\frac{a_1+\sqrt{a_1^2-4 a_2(a_0-\beta^*)}}{2a_2}, \label{energia}
\end{equation}
and from here other thermodynamic properties can be obtained, as the isochoric heat capacity,
\begin{equation}
C_v(u^*)= -\frac{\beta^2}{a_1 + 2a_2u^*}. \label{heat}
\end{equation}
Combining Eqs.~(\ref{formulaAs}) and (\ref{energia}),
the HTE coefficients can be evaluated as function of the fitting parameters $a_i$
\begin{subequations}
\label{A1-A6}
\begin{equation}
A_1= -\frac{a_1+\sqrt{a_1^2-4 a_2a_0}}{2a_2}, \label{coefA1}
\end{equation}
\begin{equation}
A_2= -\frac{1}{2\sqrt{a_1^2-4 a_2a_0}}, \label{coefA2}
\end{equation}
\begin{equation}
A_3= \frac{a_2}{3(a_1^2-4 a_2a_0)^{3/2}}, \label{coefA3}
\end{equation}
\begin{equation}
A_4= -\frac{a^2_2}{2(a_1^2-4 a_2a_0)^{5/2}}, \label{coefA4}
\end{equation}
\begin{equation}
A_5= \frac{a^3_2}{(a_1^2-4 a_2a_0)^{7/2}}, \label{coefA5}
\end{equation}
\begin{equation}
A_6= -\frac{7a^4_2}{3(a_1^2-4 a_2a_0)^{9/2}}, \label{coefA6}
\end{equation}
\end{subequations}
and so on. 

It is useful to realize that the MCE approach offers a simple and straightforward way of extracting
information of the perturbation terms by considering $\beta$ as a function of $u^*$.
From equations (18a)-(18f) we can obtain alternative representations of the perturbation terms,
\begin{equation}
A_2 = \frac{1}{2a_1 + 4a_2A_1},
\end{equation}
that corresponds to 
\begin{equation}
A_2 = -0.5\lim_{\beta^*\to 0} C_v(u^*)/\beta^2,
\end{equation}
and
\begin{equation}
A_3 = -\frac{a_2}{3}(2A_2)^3,
\end{equation}
\begin{equation}
A_4 = \frac{a_2^2}{2}(2A_2)^5,
\end{equation}
\begin{equation}
A_5 = -a_2^3(2A_2)^7,
\end{equation}
\begin{equation}
A_6 = \frac{7}{3}a_2^4(2A_2)^9.
\end{equation}
From these expressions, it is clear that the fluctuation terms that are given by
the SW perturbation coefficients of order $\ge 2$ are basically determined by the
high-temperature isochoric heat capacity, $C_v$, and $a_2 = \partial^2 \beta/\partial u^{*2}$. 
Although previous studies based on the Barker and Henderson perturbation theory modeled
$A_2$ and higher-order terms as functions of the isothermal compressibility
 approximation~\cite{ben:mp89,gilv1993,gilv1996},
{\it i.e} the fluctuation of the number of hard-spheres particles,
the MCE method presented here indicates that it could be more
appropriate to use the isochoric heat capacity.
 
\begin{figure*}
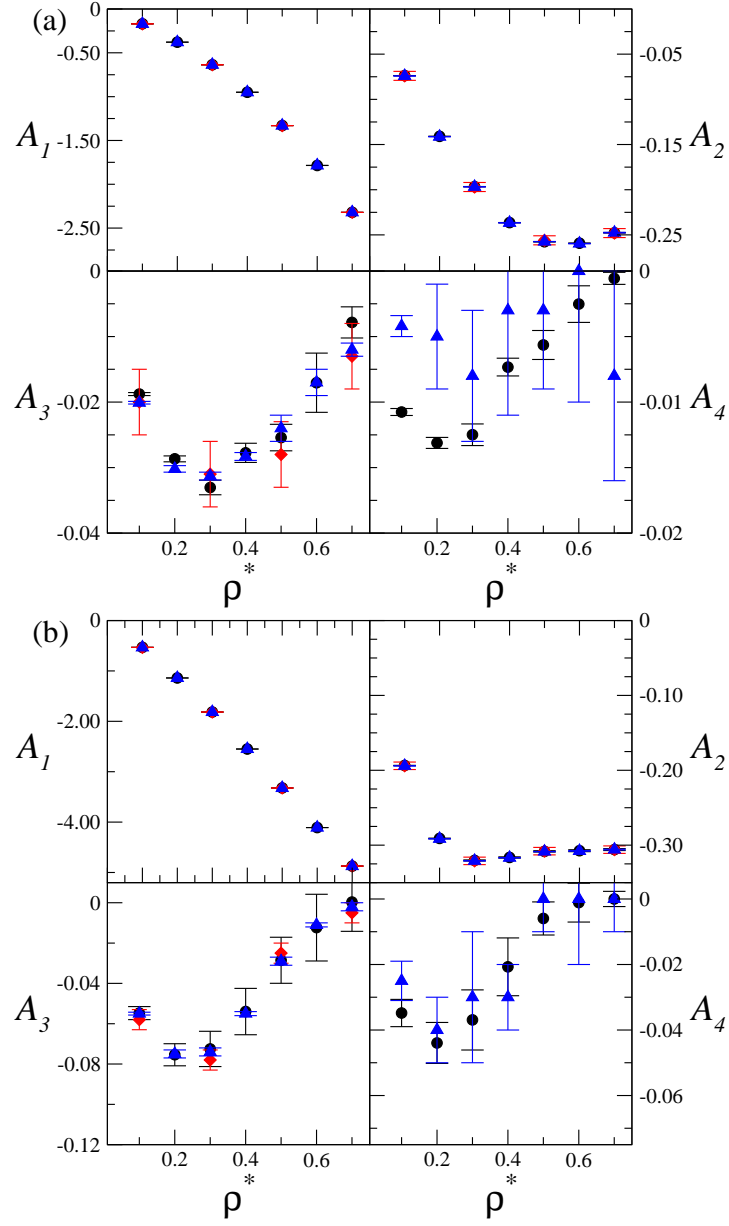

\begin{center}
\includegraphics[width=9.5cm,clip]{fig_03a.eps}
\includegraphics[width=9.5cm,clip]{fig_03b.eps}
\caption{\label{lambda1.5} (Color on line)
First HTE coefficients $A_n$ for (a) $\lambda=1.2$ and (b) $\lambda=1.5$ as function of $\rho^*$.
Black circles are results obtained with Equations~(\ref{A1-A6}), red diamonds are NVT-MC simulations
and blue triangles are values taken from Ref.~\cite{Zhou2013}.
}
\end{center}
\end{figure*}

Numerical simulations were implemented with $N\times 10^6$ particle displacement attempts
and 30 to 50 different independent runs for every set of parameters.
In Fig.~\ref{lambda1.5} results are presented
for the first four coefficients and for $\lambda=1.2$ and $\lambda=1.5 $.
NVT MC simulation results for the first three coefficients
are also included,  
these last results were obtained using
864 particles, with $2.5\times 10^5$ cycles required for thermalization and another $5.0\times 10^5$ cycles
for averaging. For comparison we include results from Zhou and Solana~\cite{Zhou2013}.
In all the cases we observe that the MCE method is in very good agreement
with the NVT and external data. Typical values of 
the absolute average relative deviation 
between the calculations with MCE and NVT MC methods were 0.11 and
0.04 for the first perturbation term with ranges $\lambda = 1.2$
and $\lambda = 1.5$, respectively. For the cases for $A_2$ and $A_3$
these values increase for the same systems: 0.17 and 0.12 for $A_2$,
8.6 and 19.3 for $A_3$, for 
$\lambda = 1.2$ and $\lambda = 1.5$, respectively.

The HTE coefficients, up to $A_6$ are listed in Table~\ref{todosAs} for all the SW ranges $\lambda$
considered in this work. 

\FloatBarrier
\begin{center}
\begin{table*}[ht!]
\vspace{-1.0cm}
\caption{\label{todosAs}
\small HTE Coefficients for all $\lambda$ ranges up to $A_6$.
}
\tiny
\begin{tabular}{ccccccccccccc}
\hline
\hline
$\rho^*$ & & $A_1$ & & $A_2$ & & $A_3$ & & $A_4$ & & $A_5$ & & $A_6$ \\
\hline
$\lambda=1.1$ \\
0.10 & &  -0.07753(97) & &  -0.0359(21) & &  -0.0087(22) & & -0.0047(26) & & -0.0034(29) & & -0.0029(33) \\
0.20 & &  -0.17686(98) & &  -0.0762(26) & &  -0.019(13) & &  -0.010(15) & & -0.008(17) & & -0.007(20) \\
0.30 & &  -0.3036(11) & &   -0.1189(32) & &  -0.025(29) & &  -0.012(30) & & -0.008(32) & & -0.006(37) \\
0.40 & &  -0.4652(13) & &   -0.1624(62) & &  -0.027(43) & &  -0.010(33) & & -0.005(25) & & -0.003(21) \\
0.50 & &  -0.6715(16) & &   -0.2045(92) & &  -0.027(92) & &  -0.008(61) & & -0.003(45) & & -0.001(37) \\
0.60 & &  -0.9350(18) & &   -0.241(11) & &   -0.02(13) & &   -0.005(66) & & -0.001(43) & & -0.000(33)\\
0.70 & &  -1.270(13)& &   -0.269(12) & &     -0.01(13) & &   -0.001(40) & & -0.000(19) & & -0.000(11) \\
$\lambda=1.2$ \\
0.10 & &  -0.16893(82) & &  -0.0737(25) & &  -0.019(11) & & -0.011(12) & &  -0.008(14) & & -0.007(17) \\
0.20 & &  -0.37858(75) & &  -0.1409(29) & &  -0.029(22) & & -0.013(20) & &  -0.008(19) & & -0.006(18) \\
0.30 & &  -0.6361(12) & &  -0.1966(55) & &  -0.033(56) & & -0.012(42) & &  -0.006(32) & & -0.004(27) \\
0.40 & &  -0.9503(11) & &  -0.2364(61) & &  -0.028(74) & & -0.007(34) & &  -0.003(18) & & -0.001(10) \\
0.50 & &  -1.3305(17) & &  -0.2576(86) & &  -0.02(11) & & -0.005(53) & &  -0.001(33) & & -0.000(25) \\
0.60 & &  -1.7851(20) & &  -0.259(15) & &  -0.02(22) & & -0.003(66) & &  -0.001(42) & & -0.000(28) \\
0.70 & &  -2.3203(17) & &  -0.248(12) & &  -0.00(11) & & -0.000(21) & &  -0.0001(71) & & -0.0000(27) \\
$\lambda=1.3$ \\
0.10 & &  -0.27460(90) & &  -0.1125(37) & &  -0.030(28) & & -0.018(33) & & -0.014(40)& & -0.013(51) \\
0.20 & &  -0.60525(75) & &  -0.1964(40) & &  -0.039(48) & & -0.017(43) & & -0.010(40)& & -0.007(39) \\
0.30 & &  -0.99738(84) & &  -0.2482(54) & &  -0.041(37) & & -0.015(28) & & -0.008(21)& & -0.004(16) \\
0.40 & &  -1.4563(15) & &  -0.2695(76) & &  -0.036(77) & & -0.011(44) & & -0.004(27)& & -0.002(18) \\
0.50 & &  -1.9842(18) & &  -0.267(10) & &  -0.033(50) & & -0.009(28) & & -0.003(16)& & -0.0014(99) \\
0.60 & &  -2.5782(20) & &  -0.255(16) & &  -0.02(11) & & -0.004(54) & & -0.001(33)& & -0.000(23) \\
0.70 & &  -3.2283(80) & &  -0.245(12) & &   0.012(78) & & -0.001(19) & & -0.0002(60)& & -0.0000(22) \\
$\lambda=1.4$ \\
0.10 & &  -0.39549(77) & &  -0.1527(41) & &  -0.040(36) & & -0.023(43) & & -0.018(51) & & -0.017(63) \\
0.20 & &  -0.85818(84) & &  -0.2454(52) & &  -0.053(30) & & -0.026(30) & & -0.017(29) & & -0.013(30) \\
0.30 & &  -1.3894(11) & &  -0.2864(60) & &  -0.055(68) & & -0.023(59)  & & -0.013(51) & & -0.009(47) \\
0.40 & &  -1.9871(17) & &  -0.292(18) & &  -0.05(13) & & -0.017(90)    & & -0.008(66) & & -0.005(53) \\
0.50 & &  -2.6437(26) & &  -0.283(17) & &  -0.03(18) & & -0.01(11)     & & -0.003(87) & & -0.001(79) \\
0.60 & &  -3.3432(49) & &  -0.279(37) & &  -0.02(26) & & -0.002(93)    & & -0.000(65) & & -0.000(48) \\
0.70 & &  -4.0611(25) & &  -0.278(10) & &   0.01(11) & & -0.001(20)   & & -0.0001(64) & & -0.000(23) \\
$\lambda=1.5$ \\
0.10 & &  -0.53247(43) & &  -0.1936(22) & &  -0.055(32) & & -0.035(42) & & -0.030(54)& & -0.029(73) \\
0.20 & &  -1.13921(70) & &  -0.2910(41) & &  -0.075(55) & & -0.044(62) & & -0.034(72)& & -0.031(86) \\
0.30 & &  -1.8150(12) & &  -0.3204(71) & &  -0.073(87) & & -0.037(92) & & -0.025(98)& & -0.02(11) \\
0.40 & &  -2.5489(15) & &  -0.3164(92) & &  -0.05(11) & & -0.020(88) & & -0.010(72)& & -0.006(64) \\
0.50 & &  -3.3224(16) & &  -0.308(10) & &  -0.03(12) & & -0.006(53) & & -0.002(28)& & -0.000(17) \\
0.60 & &  -4.1085(21) & &  -0.307(12) & &  -0.01(18) & & -0.001(55) & & -0.000(37)& & -0.000(32) \\
0.70 & &  -4.8724(23) & &  -0.306(13) & &   0.00(14) & & -0.000(24) & & -0.0000(84)& & -0.0000(29) \\
$\lambda=1.6$ \\
0.10 & &  -0.68671(81) & &  -0.2368(52) & &  -0.073(52) & & -0.051(74) & & -0.05(10)& &   -0.05(15) \\
0.20 & &  -1.4506(11) & &  -0.3356(59) & &  -0.101(53) & & -0.068(72) & &  -0.061(98)& &  -0.06(14) \\
0.30 & &  -2.2786(14) & &  -0.3550(75) & &  -0.087(99) & & -0.05(11) & &   -0.04(12)& &   -0.03(15) \\
0.40 & &  -3.1503(14) & &  -0.3468(93) & &  -0.054(94) & & -0.019(68) & &  -0.009(51)& &  -0.005(40) \\
0.50 & &  -4.0379(20) & &  -0.342(14) & &  -0.02(12) & & -0.002(35) & &    -0.000(13)& &  -0.0001(54) \\
0.60 & &  -4.9063(25) & &  -0.340(12) & &  0.00(12) & & -0.000(14) & &      0.0000(39)& & -0.00000(99) \\
0.70 & &  -5.7174(26) & &  -0.331(14) & &   0.01(12) & & -0.000(16) & &     0.0000(45)& & -0.0000(12) \\
$\lambda=1.7$ \\
0.10 & &  -0.85948(87) & &  -0.2828(56) & &  -0.098(59) & & -0.077(94) & & -0.08(15)& &   -0.10(25) \\
0.20 & &  -1.7953(11) & &  -0.3821(86) & &  -0.131(63) & &  -0.100(96) & & -0.10(15)& &   -0.12(23) \\
0.30 & &  -2.7860(11) & &  -0.3930(95) & &  -0.099(74) & &  -0.056(85) & & -0.042(99)& &  -0.04(12) \\
0.40 & &  -3.8031(19) & &  -0.383(10) & &  -0.04(12) & &    -0.011(60) & & -0.004(33)& &  -0.001(20) \\
0.50 & &  -4.8118(21) & &  -0.379(14) & &  -0.00(13) & &    -0.000(13) & & -0.000(03)& &  -0.00000(85) \\
0.60 & &  -5.7751(24) & &  -0.378(15) & &  -0.01(20) & &    -0.000(31) & &  0.000(12)& &  -0.0000(45) \\
0.70 & &  -6.6587(35) & &  -0.375(16) & &   0.01(17) & &    -0.001(26) & &  0.0001(91)& & -0.0000(33) \\
$\lambda=1.8$ \\
0.10 & &  -1.0522(11) & &  -0.3320(50) & &  -0.130(36) & & -0.114(64) & & -0.13(11)& &   -0.18(21) \\
0.20 & &  -2.1769(11) & &  -0.4316(76) & &  -0.170(81) & & -0.15(14) & &  -0.18(25)& &   -0.24(47) \\
0.30 & &  -3.3452(10) & &  -0.436(11) & &   -0.117(86) & & -0.07(10) & &  -0.06(13)& &   -0.05(16) \\
0.40 & &  -4.5225(15) & &  -0.425(12) & &   -0.05(11) & &  -0.012(53) & & -0.004(27)& &  -0.002(15) \\
0.50 & &  -5.6725(26) & &  -0.422(13) & &   -0.01(17) & &  -0.000(18) & & -0.0000(47)& & -0.0000(11) \\
0.60 & &  -6.7632(25) & &  -0.426(14) & &    0.00(18) & &  -0.000(31) & &  0.000(12)& &  -0.0000(48) \\
0.70 & &  -7.7753(32) & &  -0.437(18) & &   -0.01(16) & &  -0.000(20) & & -0.0000(52)& & -0.0000(14) \\
\hline
\hline
\end{tabular}
\end{table*}
\end{center}
\FloatBarrier

\subsection{Thermodynamic of SW fluids}

The MCE results for the HTE coefficients are required to generate the equation of state
for the monomeric SW fluid as well as for SW chain molecules in the context of the
SAFT-VR approach~\cite{gil:jcp97}. In this section we describe simple functional expressions 
to represent the MCE values, obtained by fitting appropriate analytical expressions to the simulated
values of the coefficients.

For $A_1$ the data can be represented by a third order polynomial without an independent term
\begin{equation}
A_1(\rho^*)=b_0 {\rho^*}+b_1 {\rho^*}^2+ b_2 {\rho^*}^3. \label{final-A1},
\end{equation}
whereas $A_2$ can be approximated by
\begin{equation}
A_2(\rho^*)=g_2\rho^* e^{-h_2 {\rho^*}^2} +f_2 \tanh(c_2 {\rho^*}), \label{final-A2}
\end{equation}
except for $\lambda=1.1$, where the relation is the same that the one for $A_1$.
In the same way, similar expressions can be found for the higher-order coefficients,
\begin{equation}
A_i(\rho^*)=g_i\rho^* e^{-h_i {\rho^*}^2},~~~\mbox{for}~i\ge3. \label{final-Ai}
\end{equation}

\begin{figure*}
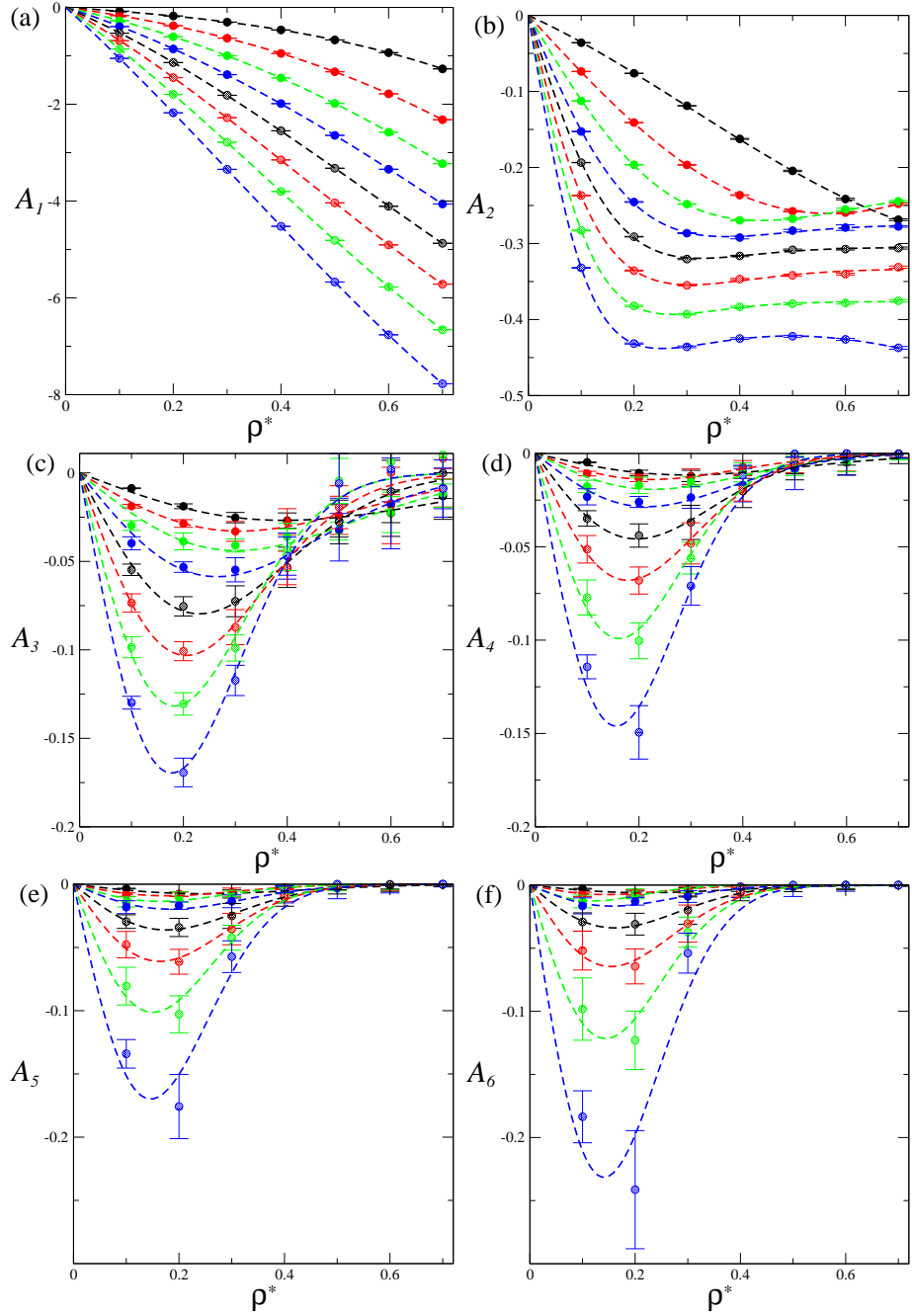

\begin{center}
\includegraphics[width=6.00cm,clip]{fig_04a.eps}
\includegraphics[width=6.00cm,clip]{fig_04b.eps} \\
\includegraphics[width=6.00cm,clip]{fig_04c.eps}
\includegraphics[width=6.00cm,clip]{fig_04d.eps} \\
\includegraphics[width=6.00cm,clip]{fig_04e.eps}
\includegraphics[width=6.00cm,clip]{fig_04f.eps}
\caption{\label{todos-ajustes} (Color on line)
HTE coefficients $A_n$ for SW ranges $\lambda$ considered in this work as function of $\rho^*$
for (a) $A_1$, (b) $A_2$, (c) $A_3$,
(d) $A_4$, (e) $A_5$ and (f) $A_5$.
In all the cases the MCE simulations are depicted as solid circles
and the dashed lines represent the fitted data according to 
relations~(\ref{final-A1}), (\ref{final-A2}) and (\ref{final-Ai}).
}
\end{center}
\end{figure*}

In Figure~\ref{todos-ajustes} results are given for the first six perturbation terms
as obtained from MCE simulations and parametrical representation.
The summary of the fitting parameters are listed in Tables \ref{TablasA1}-\ref{TablasAi}. 

\begin{table}[ht]
\caption{\label{TablasA1} Fitting parameters with Eq.~(\ref{final-A1}).
}
\begin{tabular}{ccccccccccc}
\hline
\hline
$\lambda$ & & $b_0$ & & $b_1$ & & $b_2$\\
\hline
1.10 & &  -0.74484 & &  -0.382173 & &  -1.63313 \\
1.20 & &  -1.51817 & &   -1.58568 & &   -1.40168 \\
1.30 & &  -2.40494 & &   -3.04343 & &   -0.16285 \\
1.40 & &  -3.46854 & &   -4.36238 & &   1.46006  \\
1.50 & &  -4.74229 & &   -5.34853 & &   3.10155 \\
1.60 & &  -6.24987 & &   -5.88535 & &   4.48369  \\
1.70 & &  -8.01193 & &   -5.8906   & &   5.34814 \\   
1.80 & &  -10.0523 & &   -5.27555 & &   5.38749 \\   
\hline
\hline
\end{tabular}
\end{table}

With the expressions given for the perturbation terms, and using the Carnahan-Starling equation of state for
the HS reference system, i.e.,
\begin{equation}
\frac{A_0^E}{NkT} = \frac{4\eta^2-3\eta}{(1-\eta)^2},
\end{equation}
where $\eta = \pi\rho\sigma^3/6$ is the packing fraction of the fluid,
then it is possible to predict the phase behavior of the SW fluid. For example, in Figure 5 we present
results for the liquid-vapor equilibria obtained when perturbation terms are added sequentially. 
Results are compared with Gibbs-Ensemble MC simulations ~\cite{swgibbs1992}. We can observe
that the higher the order of the perturbation term included, the lower the critical point obtained. However,
it seems that the behavior of the coexistence curves is very sensitive to
the parametrization proposed for $A_5$ and $A_6$, and that using the equation of state up to fourth order
gives the best representation of the shape of this curve.  
Since the proper modeling of the coexistence zone near the critical region
requires to include more robust approaches based on the renormalization method,
as discussed previously for the SAFT-VR method \cite{mccabe04,galindo11},
we can expect that the MCE-SW thermodynamic modeling 
could also be combined with those approaches 
in order to give a more accurate prediction of the critical properties of SW fluids. 

\begin{table}[ht!]
\caption{\label{TablasA2} Fitting parameters with Eq.~(\ref{final-A2}) for $\lambda \ge 1.2$. For $\lambda=1.1$ the
data are fitted with $A_2= -0.315714 {\rho^*} -0.405783 {\rho^*}^2 + 0.4394 {\rho^*}^3$.
}
\begin{tabular}{ccccccccccc}
\hline
\hline
$\lambda$ & & $g_2$ & & $h_2$ & & $f_2$ & & $c_2$ \\
\hline
1.20 & &  -0.754123 & &  1.54113 & &  0.00000 & &  0.00000 \\
1.30 & &  -0.400299 & &  4.81518 & &  -0.221092 & &  3.42486 \\
1.40 & &  -0.311957 & &  9.62637 & &  -0.275382 & &  4.80383 \\
1.50 & &  -0.301985 & &  12.7536 & &  -0.305239 & &  6.10676 \\
1.60 & &  -0.242938 & &  10.8684 & &  -0.333435& &  7.67705 \\
1.70 & &  -0.310091 & &  16.5368 & &  -0.376464 & &  8.31355 \\
1.80 & &  -0.385677 & &  1.98941 & &   0.539128 & &  8.40071\\
\hline
\hline
\end{tabular}
\end{table}

\begin{table*}[hb!]
\centering
\caption{\label{TablasAi} Fitting parameters with Eq.~(\ref{final-Ai}) for $i>2$.
}
\begin{tabular}{cc|ccc|cccc|cccc|ccccccccccc}
\hline
\hline
$\lambda$ & & $g_3$ & & $h_3$ & & $g_4$ & & $h_4$ & & $g_5$ & & $h_5$ & & $g_6$ & & $h_6$\\
\hline
1.10 & &  -0.112095 & & 3.21962  & &  -0.0649142 & &  5.84165 & &  -0.0504391 & & 8.30714 & & -0.0456578  & & 10.5817\\
1.20 & &  -0.175887 & &  5.26299 & &  -0.103055 & &  10.0468 & &  -0.085633 & & 16.1883 & & -0.0847178 & & 23.5119\\
1.30 & &  -0.239451 & &  5.46669 & &  -0.139246 & &  9.65685 & &  -0.136693 & & 18.7422 & & -0.170437 && 34.0676\\
1.40 & &  -0.364466 & &  7.11089 & &  -0.218461 & &  10.8009 & &  -0.177024 & & 14.9085 & & -0.175455 & & 20.3773\\
1.50 & &  -0.568689 & &  9.38497 & &  -0.385468 & &  13.0504 & &  -0.338921 & & 16.1663 & & -0.345938 & & 19.133 \\
1.60 & &  -0.828532 & &  11.8677 & &  -0.627884 & &  15.717 & & -0.615139 & & 18.6941 & & -0.69147 & & 21.1024 \\
1.70 & &  -1.18847   & &  14.9645 & &  -1.01203 & &  19.1651& & -1.11077 & & 22.1139 & & -1.39728 & & 24.2925\\
1.80 & &  -1.57846   & &  15.9349 & &  -1.54308 & &  20.5372 & & -1.91277 & & 23.335 & & -2.70865 & & 25.1763\\
\hline
\hline
\end{tabular}
\end{table*}

\begin{figure}
\begin{center}
\includegraphics[width=8.00cm]{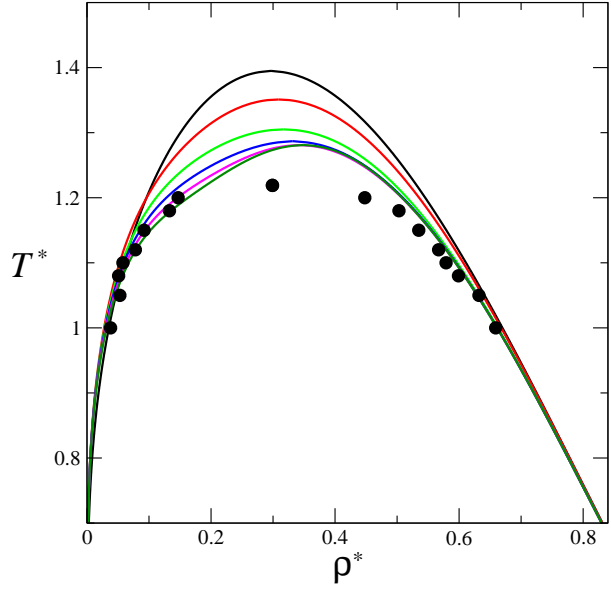}
\caption{\label{liquid-vapor} 
(Color online) Liquid-vapor equilibria for the SW fluid of range $\lambda = 1.5$
using sequential perturbation terms
from $A_1$ up to $A_6$.
 Gibbs Ensemble Monte Carlo computer simulations~\cite{swgibbs1992} are included for comparison.
}
\end{center}
\end{figure}
The convergence rate of the MCE perturbation terms can also been determined through the
isochoric heat capacity, as shown in Figure 6 for a SW fluid with range $\lambda = 1.5$
and reduced temperature $T^* = 1.5$.
When compared with the MCE simulated values for the non-perturbed system \cite{Sastre2015},
the corresponding approximation obtained through the MCE perturbation terms indicates 
that the expansion up to sixth order gives an accurate prediction of the non-perturbed value.
In the same figure we report the predictions obtained from Eqs. (25)-(27), that clearly
are very accurate.

\begin{figure}
\begin{center}
\includegraphics[width=8.00cm]{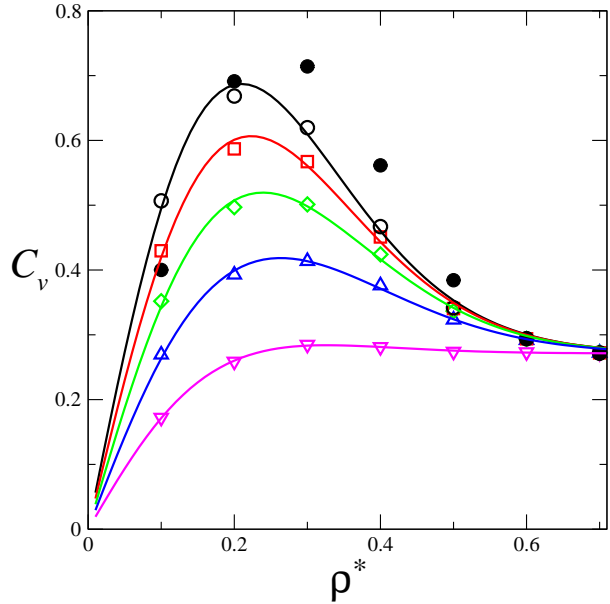}
\caption{\label{liquid-vapor} 
(Color online) Contributions to the SW fluid isochoric heat capacity
for $\lambda = 1.5$ and $T^* = 1.5$
using MCE simulated values 
from $A_2$ (purple triangles) up to $A_6$ (open circles). Solid black symbols denote
the non-perturbative MCE computer simulation values. The lines correspond to
the predictions given by equations (25)-(27) 
}
\end{center}
\end{figure}

\section{Conclusions}

We have implemented a new protocol based on the MCE method for the
evaluation of the HTE perturbation terms of the SW fluid.
This method has the advantage of allowing the evaluation with great accuracy
of higher order coefficients in the high-temperature expansion for discrete-potential systems.
Expressions for these terms have been obtained, either in parametric form as 
parabolic coefficients of fittings to the  MCE simulated values, or as analytical functions
on density and temperature that reproduce these values,
that are used to determine the equation of state of the system.
An interesting result that comes out from the MCE approach is that the SW perturbation terms
of order higher than 1 can be expressed in terms of the high-temperature value of the
isochoric heat capacity, instead of the HS isothermal compressibility as occurs in the NVT
modeling, 
suggesting an alternative route to model the equation of state for SW fluids. 
Results have been presented for the SW liquid-vapor coexistence curve and for the isochoric heat
capacity. In both cases the convergence of the perturbation expansion has been determined
from the MCE simulated values. 

The MCE is a promising approach that could help to have a better
understanding of the role of higher-order perturbation terms in equations of state used to
study the phase diagram of molecular fluids, following the same lines of research
presented by Westen and Gross~\cite{Gross17} recently developed for the HTE modeling
of continuos-potential systems 
such as the Lennard-Jones fluid.

\section*{Acknowledgments}
FS acknowledges the support by Universidad de Guanajuato (M\'exico)
Proyecto DAIP 879/2016.

\end{document}